\documentclass{ws-ijmpcs}

\usepackage{graphicx}

\newcommand{\Rslash}{\kern 0.2 em R \kern -0.56 em \raisebox{0.3ex}{/}}

\def\trimmarks{}

\begin{document}

\markboth{Marco Radici}
{Detecting correlated di-hadron pairs}

%
\catchline{}{}{}{}{}
%

\title{DETECTING CORRELATED DI-HADRON PAIRS: \\ ABOUT THE EXTRACTION OF TRANSVERSITY AND BEYOND}

\author{MARCO RADICI\footnote{Speaker}}

\address{INFN - Sezione di Pavia, via Bassi 6\\ Pavia, I-27100, Italy\\
marco.radici@pv.infn.it}

\author{ALESSANDRO BACCHETTA}

\address{University of Pavia and INFN - Sezione di Pavia, via Bassi 6\\ Pavia, I-27100, Italy\\
alessandro.bacchetta@unipv.it}

\author{A. COURTOY}

\address{IFPA, AGO Department, University of Li\`ege and INFN - LNF Frascati\\ Li\`ege, Belgium\\
aurore.courtoy@ulg.ac.be}

\maketitle


\begin{abstract}
We summarize the latest achievements about the extraction of the transversity parton distribution and proton tensor charge based on an analysis of pion-pair production in deep-inelastic scattering off transversely polarized targets. Recently released data for proton and deuteron targets by HERMES and COMPASS allow for a flavor separation of the valence components of transversity. At variance with the Collins effect, this extraction is performed in the framework of collinear factorization and relies on di-hadron fragmentation functions. The latter have been taken from the first recent analysis of the semi-inclusive production of two pion pairs in back-to-back jets in $e^+ e^-$ annihilation. We also comment on the possibility of isolating new azimuthally asymmetric correlations of opposite pion pairs, which could arise when a fragmenting quark crosses parity-odd domains localized in Minkowski space-time and induced by the topologically nontrivial QCD background (the so-called $\theta$ vacuum). 

\keywords{QCD Phenomenology; Deep-Inelastic Scattering; Parity violation.}
\end{abstract}

\ccode{PACS numbers: 13.87.Fh, 13.88.+e, 13.66.Bc, 11.30.Er}

\section{Introduction}
\label{intro}	

In a parton model picture, parton distribution functions (PDFs) describe combinations of number densities of quarks and gluons in a fast-moving hadron and are a crucial ingredient for our understandings of high-energy experiments involving hadrons.  At leading twist, the quark structure of spin-half hadrons is described by three PDFs: the unpolarized distribution $f_1(x)$, the longitudinal polarization (helicity) distribution $g_1(x)$, and the transverse
polarization (transversity) distribution $h_1(x)$. From the phenomenological point of view, $h_1(x)$ is the least known one because, being chiral-odd, it can be measured only in processes with two hadrons in the initial state 
({\it e.g.} proton-proton collision) or one hadron in the initial state and at least one hadron in the final state ({\it e.g.} semi-inclusive DIS - SIDIS).  

Combining data on polarized single-hadron SIDIS, together with data on almost back-to-back emission of two hadrons in $e^+ e^-$ annihilations, the transversity distribution was extracted for the first time by the Torino 
group~\cite{Anselmino:2008jk,Anselmino:2013}. The main difficulty of such analysis lies in the factorization framework used to interpret the data, since it involves Transverse Momentum Dependent PDFs (TMDs). QCD evolution of TMDs must be included to analyze SIDIS and $e^+ e^-$ data obtained at very different scales, but the computation of its effects is still under active debate. Recently, the impact of evolution on $h_1(x)$ was estimated in the so-called CSS framework~\cite{AleAlexei:2013}. 

Alternatively, the transversity distribution can be extracted in the standard framework of collinear factorization using 
data on SIDIS with two hadrons detected in the final state. In fact, $h_1(x)$ is multiplied with a specific chiral-odd 
Di-hadron Fragmentation Function (DiFF)~\cite{Collins:1994kq,Jaffe:1998hf,Radici:2001na}, which can be extracted from the corresponding $e^+ e^-$ annihilation process leading to two back-to-back pion pairs~\cite{e+e-:2003}. The collinear framework allows to keep under control the evolution equations of DiFFs~\cite{Ceccopieri:2007ip}. Using two-hadron SIDIS data on proton and deuteron targets from HERMES~\cite{DiFFHERMES} and COMPASS~\cite{Adolph:2012nw}, as well as Belle data for the process $e^+ e^- \to (\pi^+ \pi^-) (\pi^+ \pi^-) X$~\cite{Vossen:2011fk}, the transversity $h_1(x)$ was extracted for the first time in the collinear framework~\cite{Bacchetta:2011ip} and the valence components of up and down quark were 
separated~\cite{h1JHEP}. In this contribution to the proceedings, we summarize the parametrization and the error analysis both for $h_1(x)$ and for the DiFFs~\cite{noiBelle}. 

We also comment on re-analyzing the $e^+ e^- \to (\pi^+ \pi^-) (\pi^+ \pi^-) X$ process by including structures violating parity invariance (P-odd structures). P-odd DiFFs arise when the fragmenting quark crosses local (in space and time) domains where strong interactions break parity invariance. These P-odd bubbles can be induced by topologically nontrivial solutions of QCD equations of motion, which suggest that the physical vacuum is a superposition of degenerate states, the so-called $\theta$ 
vacuum~\cite{KK3}, and that topological fluctuations in this vacuum structure can be equivalently represented by a local P-odd term in the QCD Lagrangian~\cite{KK5}. P-odd DiFFs generate new azimuthally asymmetric terms in the cross section for the $e^+ e^- \to (\pi^+ \pi^-) (\pi^+ \pi^-) X$ process, similarly to what happens for the 
$e^+ e^- \to \pi^+ \pi^- X$ one~\cite{KK:2011} but in a collinear framework. Here, we write the full structure of the leading twist cross section and we try to estimate the size of the P-odd contributions. 


\section{Theoretical Framework for Di-hadron Semi-inclusive Production}
\label{theory}

We consider the process $\ell(k) + N(P) \to \ell(k') + H_1(P_1) + H_2(P_2) + X$, where $\ell$ denotes the beam lepton, $N$ the nucleon target with mass $M$ and polarization $S$, $H_1$ and $H_2$ the produced unpolarized hadrons with masses $M_1$ and $M_2$, respectively (the four-momenta are given in parentheses). We define the total $P_h = P_1 + P_2$ and relative $R = (P_1-P_2)/2$ momenta of the pair, with $P_h^2 = M_h^2 \ll Q^2=-q^2$ and $q = k - k'$ the momentum transferred. We define the azimuthal angles $\phi_R$ and $\phi_S$ as the angles of 
${\bf R}_T$ and ${\bf S}_T$, respectively, around the virtual photon direction ${\bf q}$. We also define the polar angle $\theta$ which is the angle between the direction of the back-to-back emission in the center-of-mass (cm) frame of the two hadrons, and the direction of $P_h$ in the photon-proton cm frame. Then, ${\bf R}_T = {\bf R} \sin\theta$ and 
$|{\bf R}|$ is a function of the invariant mass only~\cite{Bacchetta:2002ux}. Finally, we usually define the SIDIS invariants 
\begin{equation}
\centering
x = \frac{Q^2}{2\, P\cdot q} \; , \quad y = \frac{P \cdot q}{P \cdot k} \; , \quad z = \frac{P \cdot P_h}{P \cdot q} \equiv z_1 + z_2 \; .
\label{eq: invariants}
\end{equation}
To leading twist, the differential cross section for the two-hadron SIDIS off a transversely polarized nucleon target becomes~\cite{h1JHEP}
\begin{equation}
\frac{d\sigma}{dx \, dy\, dz\, d\phi_S\, d\phi_R\, d M_{h}^2\,d \cos{\theta}} =  \frac{\alpha^2}{x y\, Q^2}\, 
\Biggl\{ A(y) \, F_{UU}  + |\bm{S}_T|\, B(y) \, \sin(\phi_R+\phi_S)\,  F_{UT} \Biggr\} \; ,
\label{crossSIDIS}
\end{equation}
where $\alpha$ is the fine structure constant, $A(y) = 1-y+y^2/2$, $B(y) = 1-y$, and 
\begin{eqnarray} 
F_{UU} & = &x \sum_q e_q^2\, f_1^q(x; Q^2)\, D_1^q\bigl(z,\cos \theta, M_h; Q^2\bigr) \; , \nonumber \\
F_{UT} &=  &\frac{|{\bf R}| \sin \theta}{M_h}\, x\, 
\sum_q e_q^2\,  h_1^q(x; Q^2)\,H_1^{\sphericalangle\, q}\bigl(z,\cos \theta, M_h; Q^2\bigr) \; , 
\label{StructFunct}
\end{eqnarray}
with $e_q$ the fractional charge of a parton with flavor $q$. The $D_1^q$ is the DiFF describing the hadronization of an unpolarized parton with flavor $q$ into an unpolarized hadron pair. The $H_1^{\sphericalangle\, q}$ is its 
chiral-odd partner describing the same fragmentation but for a transversely polarized parton~\cite{Bianconi:1999cd}. DiFFs can be expanded in Legendre polynomials in $\cos \theta$~\cite{Bacchetta:2002ux}. After averaging over
$\cos \theta$, only the term corresponding to the unpolarized pair being created in a relative $\Delta L=0$ state survives in the $D_1$ expansion, while the interference in $|\Delta L| = 1$ survives for 
$H_1^{\sphericalangle}$~\cite{Bacchetta:2002ux}. Without ambiguity, the two terms will be identified with $D_1$ and $H_1^{\sphericalangle}$, respectively. 

Inserting the structure functions of Eq.~(\ref{StructFunct}) into the cross section~(\ref{crossSIDIS}), we get the single-spin asymmetry (SSA)~\cite{Radici:2001na,Bacchetta:2002ux,Bacchetta:2006un}
\begin{equation}
A_{{\rm SIDIS}}(x, z, M_h; Q) =  - \frac{B(y)}{A(y)} \,\frac{|\bm{R} |}{M_h} \, 
\frac{ \sum_q\, e_q^2\, h_1^q(x; Q^2)\, H_1^{\sphericalangle\, q}(z, M_h; Q^2)    } 
        { \sum_q\, e_q^2\, f_1^q(x; Q^2)\, D_{1}^q (z, M_h; Q^2) }\;  .
\label{SIDISssa}
\end{equation} 
For the specific case of production of $\pi^+ \pi^-$ pairs, isospin symmetry and charge conjugation suggest 
$D_1^q = D_1^{\bar{q}}$ and $H_1^{\sphericalangle\, q} = - H_1^{\sphericalangle\, \bar{q}}$, with $q=u,d,s$, 
with also $H_1^{\sphericalangle\, u} = - H_1^{\sphericalangle\, d}$~\cite{Bacchetta:2006un,Bacchetta:2011ip}. Moreover, from Eq.~(\ref{SIDISssa}) the $x$-dependence of transversity is more conveniently studied by integrating the $z$- and $M_h$-dependences of DiFFs. So, the actual combinations of transversity used in the SIDIS analysis are, for the proton target~\cite{h1JHEP}, 
\begin{eqnarray} 
x\, h_1^{p}(x; Q^2) &\equiv &x \, h_1^{u_v}(x; Q^2) - {\textstyle \frac{1}{4}}\, x h_1^{d_v}(x; Q^2) \nonumber \\
&= &-\frac{ A^p_{{\rm SIDIS}} (x; Q^2)  }{n_u^{\uparrow}(Q^2)}\,\frac{A(y)}{B(y)} \, \frac{9}{4} \sum_{q=u,d,s} \, e_q^2\, 
n_q (Q^2)\, x f_1^{q+\bar{q}}(x; Q^2) \; , 
\label{xh1p}
\end{eqnarray}
and, for the deuteron target, 
\begin{eqnarray} 
 x\, h_1^{D} (x; Q^2) &\equiv &x \, h_1^{u_v}(x; Q^2)+ x h_1^{d_v}(x; Q^2)   \nonumber \\
 &= &- \frac{A^D_{\text{SIDIS}}(x; Q^2)}{n_u^{\uparrow}(Q^2)} \,\frac{A(y)}{B(y)} \, 3 \sum_{q=u,d,s}\, 
 \big[ e_q^2\, n_q (Q^2) + e_{\tilde{q}}^2\, n_{\tilde{q}} (Q^2) \big] \, x f_1^{q+\bar{q}}(x; Q^2) \; , \nonumber \\
& &  \label{xh1D}
\end{eqnarray}
where $h_1^{q_v} \equiv h_1^q - h_1^{\bar{q}}$, $f_1^{q+\bar{q}} \equiv f_1^q + f_1^{\bar{q}}$, $\tilde{q}=d,u,s$ if $q=u,d,s$, respectively ({\it i.e.} it reflects isospin symmetry of strong interactions inside the deuteron), and 
\begin{eqnarray} 
n_q(Q^2) &= &\int_{z_{{\rm min}}}^{z_{{\rm max}}} \int_{M_{h\, {\rm min}}}^{M_{h\, {\rm max}}} 
dz \, dM_h \, D_1^q (z, M_h; Q^2)  \nonumber  \\
n_q^\uparrow (Q^2) &= &\int_{z_{{\rm min}}}^{z_{{\rm max}}} \int_{M_{h\, {\rm min}}}^{M_{h\, {\rm max}}} 
dz \, dM_h \, \frac{|{\bf R}|}{M_h}\, H_1^{\sphericalangle\, q}(z,M_h; Q^2) \; .
\label{DiFFnq}
\end{eqnarray} 
The latter quantities can be determined by extracting DiFFs from the $e^+ e^- \to (\pi^+ \pi^-) (\pi^+ \pi^-) X$ process. In fact, after the annihilation a quark and an antiquark are emitted back-to-back, each one fragmenting into a residual jet and a 
$(\pi^+ \pi^-)$ pair. The leading-twist cross section in collinear factorization, namely by integrating upon all transverse momenta but ${\bf R}_T$ and ${\bf \bar{R}}_T$, can be written as~\cite{noiBelle}
\begin{equation}
d\sigma = \frac{1}{4\pi^2}\, d\sigma^0 \, \bigg( 1+ \cos (\phi_R + \phi_{\bar{R}} ) \, A_{e+e-} \bigg) \; , 
\label{e+e-cross}
\end{equation}
where $d\sigma^0$ is the symmetric term corresponding to the production of unpolarized quark and antiquark, and the azimuthal angles $\phi_R$ and $\phi_{\bar{R}}$ give the orientation of the planes containing the momenta of the pion pairs with respect to the lepton plane (see Fig.1 of Ref.~\refcite{noiBelle} for more details). The azimuthally asymmetric term is generated from the production of the quark and antiquark being transversely polarized. We define the so-called Artru-Collins asymmetry~\cite{e+e-:2003}
\begin{equation}
A_{e+e-} \propto \frac{|{\bf R}_T|}{M_h} \, \frac{|{\bf \bar{R}}_T|}{\bar{M}_h} \, 
\frac{\sum_q e_q^2\, H_1^{\sphericalangle\, q}(z,M_h; Q^2)\, H_1^{\sphericalangle\, \bar{q}}(\bar{z},\bar{M}_h; Q^2)}{\sum_q e_q^2\, D_1^q(z,M_h; Q^2)\, D_1^{\bar{q}}(\bar{z},\bar{M}_h; Q^2)} \, .
\label{e+e-ssa}
\end{equation} 

Before the measurement of the Artru-Collins asymmetry by the Belle collaboration~\cite{Vossen:2011fk}, the only information available on DiFFs were coming from model calculations in the context of the spectator 
approximation~\cite{Bianconi:1999uc,Radici:2001na,Bacchetta:2006un}, which have produced solid predictions successfully compared with data for the asymmetry~(\ref{SIDISssa}) recently released by the COMPASS 
collaboration~\cite{Adolph:2012nw}.

\begin{figure}[pb]
\begin{center}
\includegraphics[width=6.3cm]{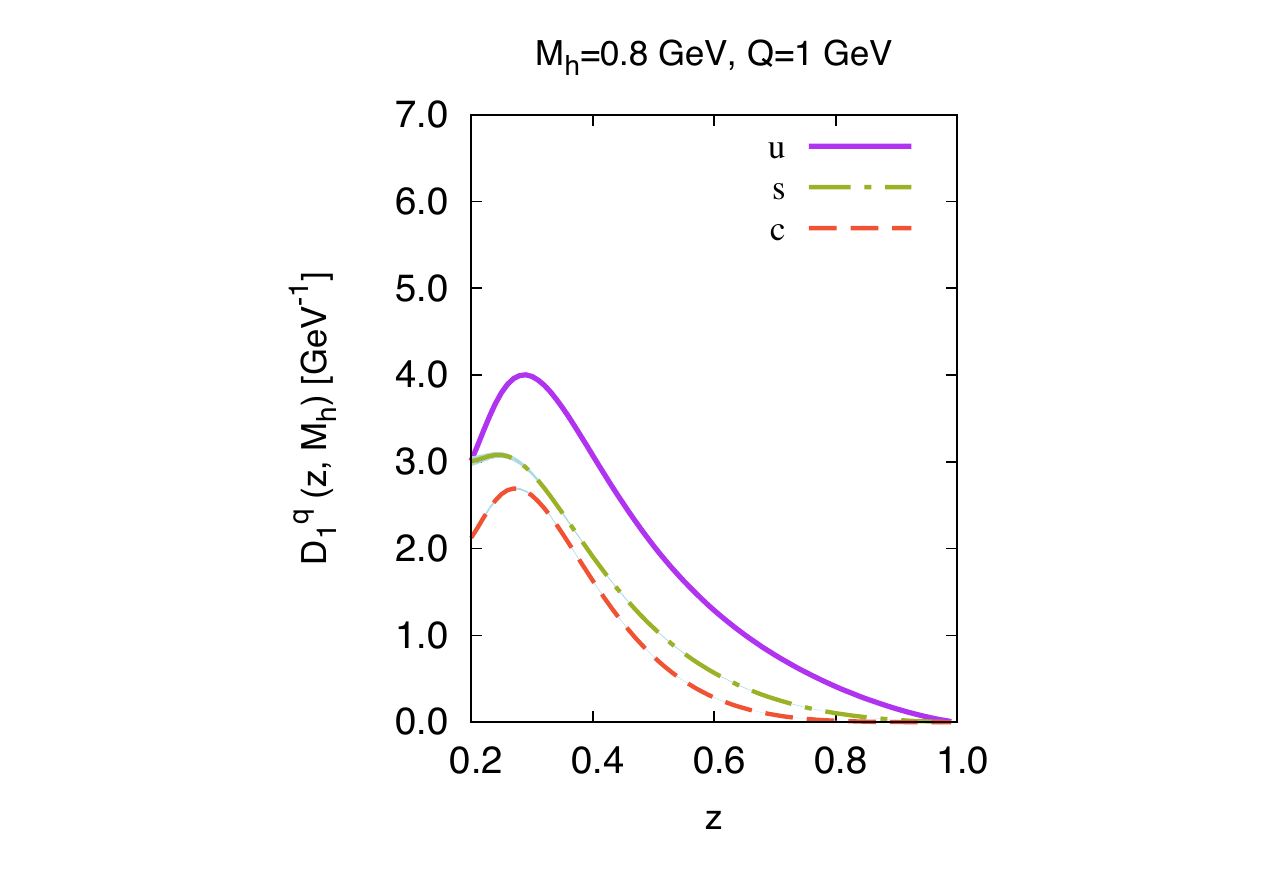}\hspace{0cm}\includegraphics[width=6.3cm]{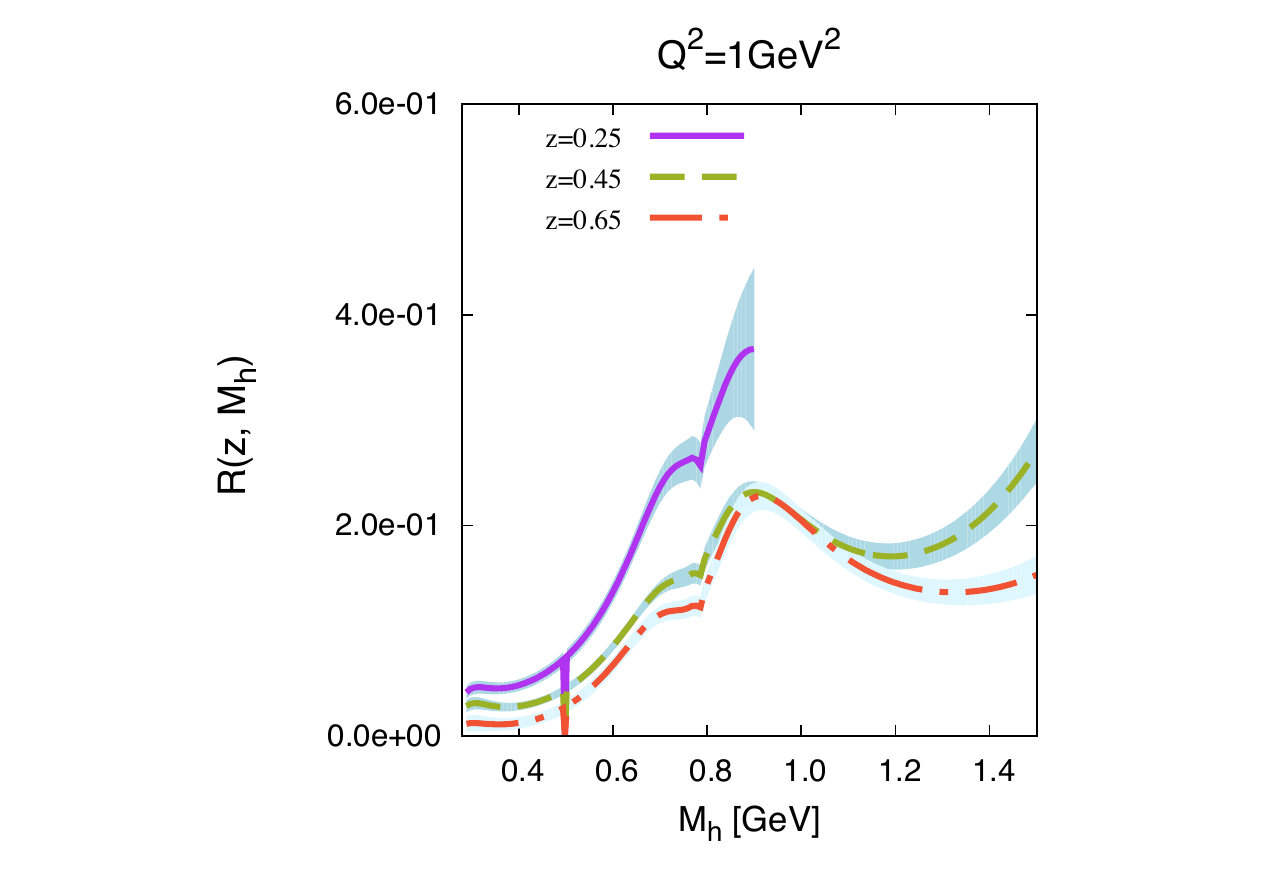}
\end{center}
\vspace*{8pt}
\caption{Left panel: $D_1^q$ as a function of $z$ at $Q_0^2=1$ GeV$^2$ and $M_h=0.8$ GeV for various flavors. Right panel: 
$(|{\bf R}| / M_h) \, (H_1^{\sphericalangle\, u} / D_1^u)$ as a function of $M_h$ at $Q_0^2=1$ 
GeV$^2$ for three different $z=0.25,\, 0.45,\, 0.65$. 
\label{D1}}
\end{figure}

Since a measurement of the unpolarized $e^+ e^-$ cross section $d\sigma^0$ is still missing, the unpolarized DiFF $D_1$ was parametrized to reproduce the two-pion yield of the PYTHIA event generator tuned to the Belle kinematics~\cite{noiBelle}. The fitting expression at the starting scale $Q_0^2=1$ GeV$^2$ was inspired by the above mentioned spectator model 
calculations~\cite{Bacchetta:2006un,Radici:2001na,Bianconi:1999uc,SIDISe+e-2009} and it contains three resonant channels (pion pair produced by $\rho$, $\omega$, and $K^0_S$ decays) and a continuum. For each channel and for each flavor 
$q= u,d,s,c$, a grid of data in $(z,M_h)$ was produced using PYTHIA for a total amount of approximately 32000 bins. Each grid was separately fitted using the corresponding parametrization of $D_1$ and evolving it to the Belle scale of $Q^2=100$ GeV$^2$. An average $\chi^2$ per degree of freedom ($\chi^2$/dof) of 1.62 was reached using in total 79 parameters (see 
Ref.~\refcite{noiBelle} for more details). In Fig.~\ref{D1}, the $D_1^q$ for $q=u(=d),s,c$ flavors at $Q_0^2=1$ GeV$^2$ and $M_h=0.8$ GeV is reported as a function of $z$ (left panel). 

Then, the chiral-odd DiFF $H_1^{\sphericalangle}$ was extracted from the Artru-Collins asymmetry by using the above mentioned isospin symmetry and charge conjugation of DiFFs and by integrating upon the hemisphere of the antiquark 
jet~\cite{noiBelle}. The experimental data for $A_{e+e-}$ are organized in a $(z,M_h)$ grid of 64 bins~\cite{Vossen:2011fk}. They were fitted starting from an expression for $H_1^{\sphericalangle\, u}$ at $Q_0^2=1$ GeV$^2$ with 9 parameters, and then evolving it to the Belle scale. The final $\chi^2$/dof was 0.57~\cite{noiBelle}. In the right panel of Fig.~\ref{D1}, the ratio 
$(|{\bf R}|/M_h) \, (H_1^{\sphericalangle\, u}/D_1^u)$ at $Q_0^2=1$ GeV$^2$ is reported as a function of $M_h$ for three different $z=0.25,\, 0.45,\, 0.65$ (for a more detailed discussion also of errors, see Ref.~\refcite{noiBelle}). 


\section{Extraction of Transversity}
\label{paramh1}

The knowledge of DiFFs in Eq.~(\ref{xh1p}) allowed us to get a glimpse of the combination 
$h_1^{u_v} - h_1^{d_v}/4$ directly from the HERMES data for $A_{{\rm SIDIS}}^p$~\cite{Bacchetta:2011ip}. Recently, the COMPASS collaboration has released new data for $A_{{\rm SIDIS}}^p$ on a proton target and for 
$A_{{\rm SIDIS}}^D$ on a deuteron target~\cite{Adolph:2012nw}. Thus, the combination of Eqs.~(\ref{xh1p}) and (\ref{xh1D}) made it possible to separately parametrize for the first time each valence flavor of the transversity 
distribution~\cite{h1JHEP}. Here, we summarize the fitting procedure and error analysis, and comment some significant results. 

\subsection{Fitting procedure and error analysis}
\label{fit}

The main theoretical constraint on transversity is Soffer's inequality~\cite{Soffer:1995ww}. If the Soffer bound is fulfilled at some initial $Q_0^2$, it will hold also at higher $Q^2 \geq Q_0^2$. We impose this condition by multiplying the functional form by the corresponding Soffer bound $\mbox{\small SB}^q(x; Q^2)$ at the starting scale $Q_0^2=1$ GeV$^2$. The Soffer bound is built from the MSTW08 set~\cite{Martin:2009iq} for $f_1$, and from the DSSV parameterization~\cite{deFlorian:2009vb} for $g_1$. Our analysis was carried out at LO in 
$\alpha_S$; the explicit form of SB$^q$ is reported in Appendix of Ref.~\refcite{h1JHEP}. The functional form for the valence transversity distributions at $Q_0^2 = 1$ GeV$^2$ reads
\begin{equation} 
x\, h_1^{q_v}(x; Q_0^2)=
\tanh \Bigl[ x^{1/2} \, \bigl( A_q+B_q\, x+ C_q\, x^2+D_q\, x^3\bigr)\Bigr]\, x \, 
\Bigl[ \mbox{\small SB}^q(x; Q_0^2)+ \mbox{\small SB}^{\bar q}(x; Q_0^2)\Bigr] \, .
\label{funct_form}
\end{equation} 
The hyperbolic tangent is such that the Soffer bound is always fulfilled. The low-$x$ behavior is determined by the $x^{1/2}$ term, which is imposed by hand to grant the integrability of Eq.~(\ref{funct_form}) and a finite tensor charge. Present fixed-target data do not allow to constrain it. The functional form is very flexible and can contain up to three nodes. Here, we show the results employing all the parameters, the so-called {\it extra-flexible scenario} (for results with other choices, see Refs.~\refcite{h1JHEP}). 

The fit and the error analysis were carried out in two different ways: using the standard Hessian method and using a Monte Carlo approach. The latter was introduced because it does not rely on the assumptions of a quadratic
dependence of $\chi^2$ and a linear expansion of the theoretical quantity around the minimum, which are prerequisites for the Hessian method. This freedom is essential whenever the minimization pushes the theoretical function towards its upper or lower bounds. The Monte Carlo approach is inspired to the work of the NNPDF collaboration~\cite{Forte:2002fg}, although our results are not based on a neural-network fit. The approach consists in creating $N$ replicas of the data points by shifting them by a Gaussian noise with the same variance as the measurement. Each replica, therefore, represents a possible outcome of an independent measurement. Then, the standard minimization procedure is applied to each replica separately (for details, see Ref.~\refcite{h1JHEP}). The number of replicas is chosen so that the mean and standard deviation of the set of replicas accurately reproduces the original data points. In our case, it turns out to be 100. The $N$ theoretical outcomes can have any distribution, not necessarily Gaussian. In which case, the $1\sigma$ confidence interval is in general different from the 68\% interval. The latter can be easily obtained by rejecting the largest and lowest 16\% of replicas for each experimental point.

\begin{figure}[pb]
\begin{center}
\includegraphics[width=6cm]{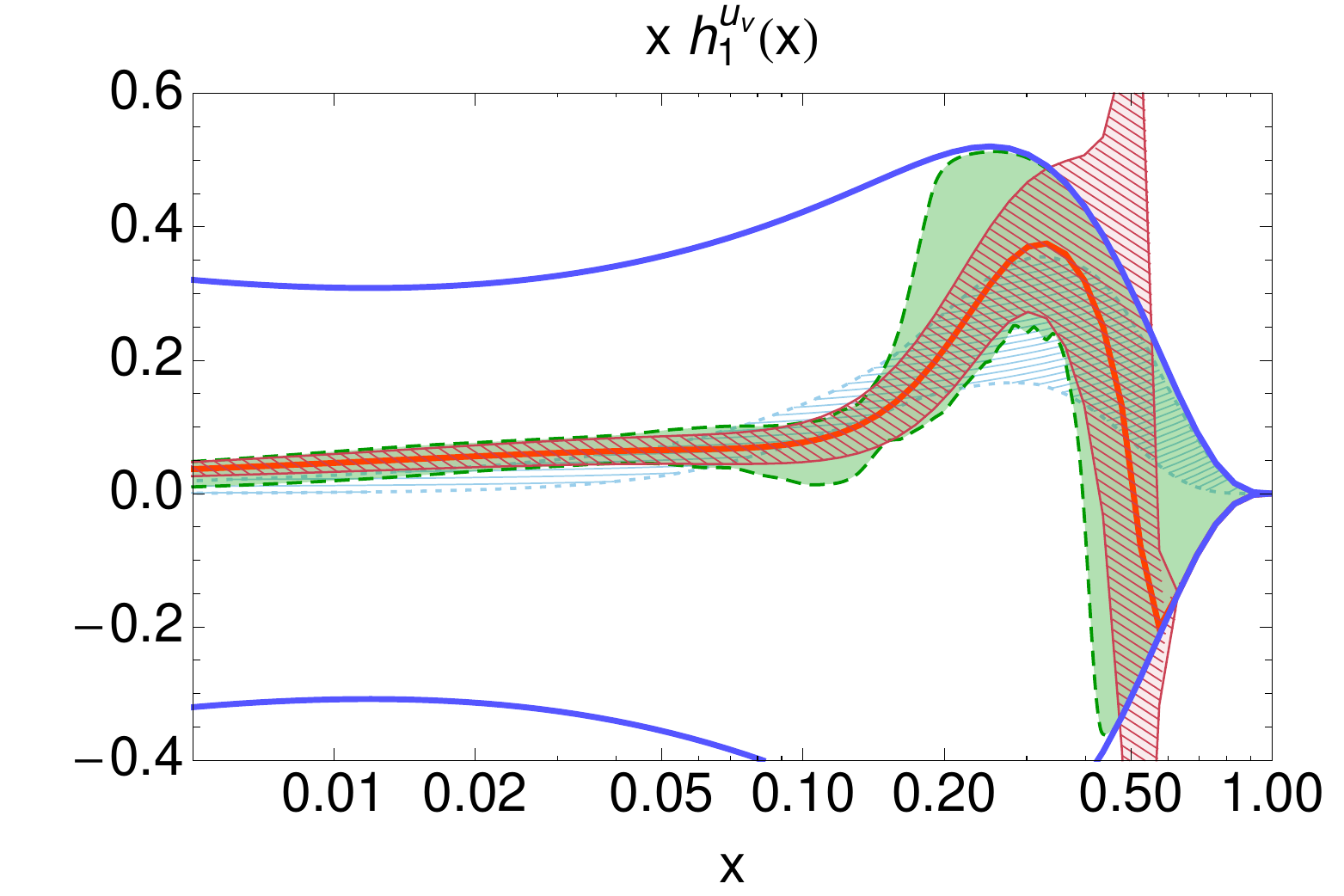}\hspace{0.5cm}\includegraphics[width=6cm]{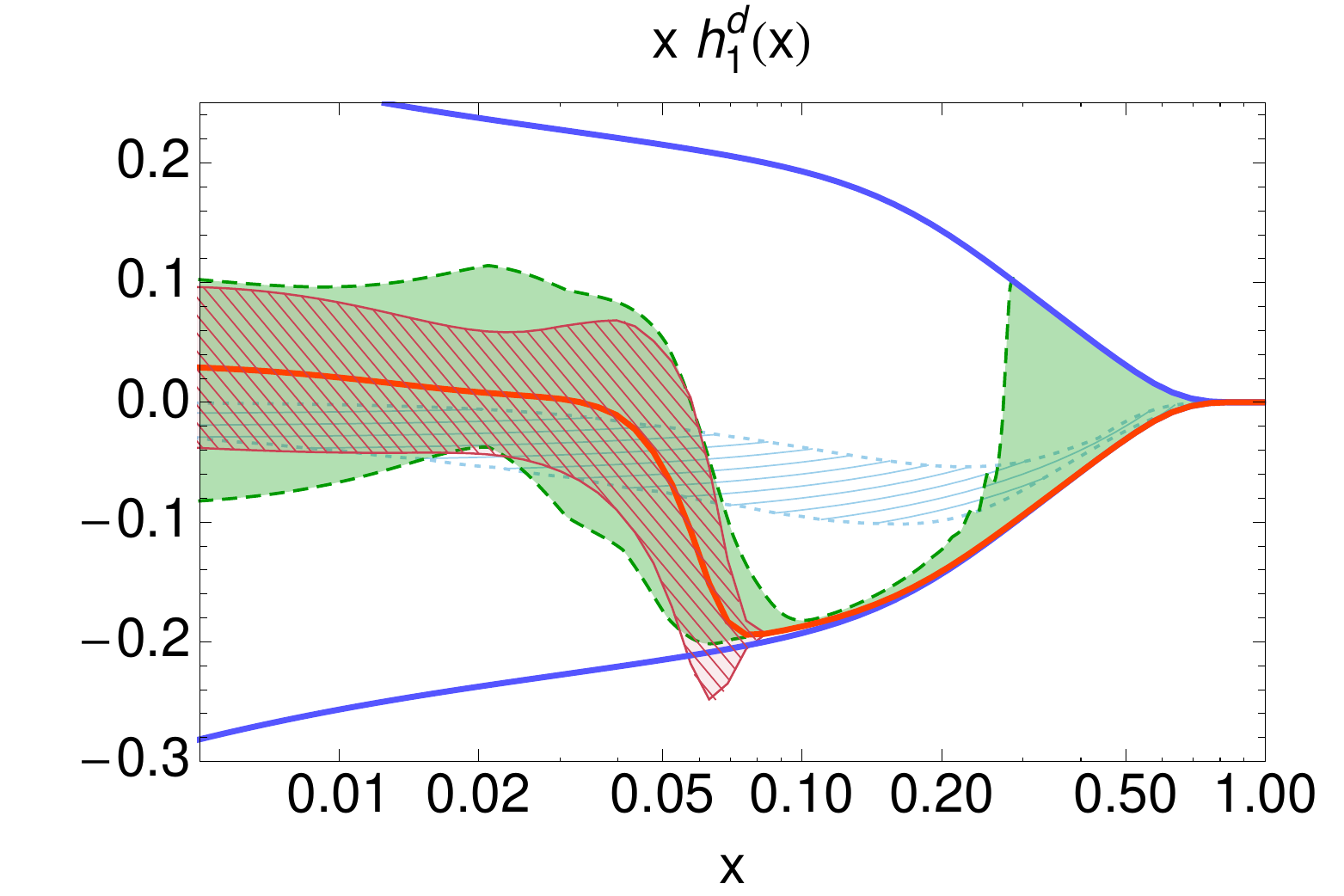}
\end{center}
\vspace*{8pt}
\caption{Left panel: the $u_v$ contribution to the transversity in the {\it extra-flexible scenario}. The standard Hessian method gives the uncertainty band with solid boundaries, the central thick solid line visualizing the central value. The band with dashed boundaries is the outcome when adopting the Monte Carlo approach. The band with short-dashed boundaries is the parametrization obtained in Ref. 1 from the Collins effect. 
The dark thick solid lines indicate the Soffer bound. All results at the scale of $Q^2=2.4$ GeV$^2$. Right panel: same notations for the $d_v$ contribution.  \label{xh1}}
\end{figure}

\subsection{Results}
\label{out}

In Fig.~\ref{xh1}, the left panel displays the $q = u_v$ contribution in Eq.~(\ref{funct_form}) for the {\it extra-flexible scenario}, while $q = d_v$ is in the right one. For each panel, the outcome with the standard Hessian method is represented by the uncertainty band with solid boundaries, the central thick solid line visualizing the central value. The partially overlapping band with dashed boundaries is the outcome when adopting the Monte Carlo approach, where the band width corresponds to the $68\%$ of all the 100 replicas, produced by rejecting the largest and lowest $16\%$ among the replicas' values in each $x$ point. For sake of comparison, each panel displays also the corresponding results in the framework of the Collins 
effect~\cite{Anselmino:2008jk}, depicted as a band with short-dashed boundaries. Since the latter was extracted at the scale $Q^2 = 2.4$ GeV$^2$, our results are properly evolved at the same scale. Finally, the dark thick solid lines indicate the Soffer bound, also evolved at the same scale. 

The uncertainty bands in the standard and Monte Carlo approaches are quite similar within the $x$-range where data exist, namely $0.0064\leq x \leq 0.28$, and are in agreement with the other extraction based on the Collins effect. Outside this range, the standard approach tends to saturate the Soffer bound, and the boundaries of the band can occasionally cross it because the assumed quadratic dependence of $\chi^2$ on the parameters around its minimum is not reliable. On the contrary, in the Monte Carlo approach each replica is built such that it never violates the Soffer bound. For $x \geq 0.4$, the replicas entirely fill the area between the upper and lower Soffer bounds, giving an explicit visualization of the realistic degree of uncertainty where there are no experimental data points. In the right panel, for $x \geq 0.1$ our results tend to saturate the lower limit of the Soffer bound because they are driven by the COMPASS deuteron data, in particular by the bins number 7 and 8. No such trend is evident in the parametrization corresponding to the single-hadron measurement of the Collins effect. As a matter of fact, this is the only source of significant discrepancy between the two extractions. This statement remains valid also for the other {\it scenarios}, indicating that this is not an artifact of the chosen functional form. 

The tensor charge, a fundamental quantity of hadrons at the same level as the vector, axial, and scalar charges, is related to the transversity by 
\begin{equation}
\delta q_v (Q^2) = \int dx \, h_1^{q_v} (x, Q^2) \; . 
\label{e:tensch}
\end{equation} 
The region of validity of our fit is restricted to the experimental data range. We can therefore give a reliable estimate for the tensor charge by truncating the integral to the interval $0.0064 \leq x \leq 0.28$. For the {\it extra-flexible scenario}, we find at $Q_0^2=1$ GeV$^2$ 
\begin{eqnarray}
&\delta u_v = 0.32 \pm 0.12 &\quad \delta d_v = -0.25 \pm 0.15 \quad \mbox{standard} \nonumber \\
&\delta u_v = 0.34 \pm 0.10 &\quad \delta d_v = -0.20 \pm 0.14 \quad \mbox{Monte Carlo} \: . 
\label{tensorvalues}
\end{eqnarray}
These results are compatible with other {\it scenarios} and, within errors, with the extraction in the framework of the single-hadron Collins effect~\cite{Anselmino:2008jk}. We tried also to extend the range of integration outside the experimental data to $0 < x \leq 1$. The result is heavily influenced by the adopted functional form, in particular by the low-$x$ exponent, and this fact reflects again the uncertainty because of missing data at very low $x$.  


\section{Parity Violation in Quark Fragmentation}
\label{PV}

QCD equations of motion possess topologically nontrivial solutions which suggest that the physical vacuum is a superposition of degenerate states differing by their topological charge. This vacuum structure is conventionally addressed as the $\theta$ vacuum~\cite{KK3}. It can be reflected in the QCD Lagrangian by introducing a term proportional to the parameter $\theta$ that globally breaks P and CP symmetries of QCD. Experimental findings indicate that $\theta$ is very small. Nevertheless, topological fluctuations in this vacuum structure can be equivalently represented by making $\theta$ varying in space and time, therefore locally breaking P and 
CP~\cite{KK5}. Consequently, during fragmentation a quark can cross local domains (P-odd bubbles) where 
P-invariance of strong interactions is broken. The most general Lorentz-invariant decomposition of the leading-twist quark-quark correlator for fragmentation in a pair of pseudoscalar mesons, depicted in 
Ref.~\refcite{Bianconi:1999cd}, can then be generalized to include terms violating parity invariance. In the collinear case, it reads
\begin{eqnarray}
\Delta (z, \zeta, {\bf R}_T) &= &\frac{1}{4\pi} \left[ D_1 (z, \zeta, {\bf R}_T) + i \frac{\Rslash_T}{M_h} 
H_1^\sphericalangle (z, \zeta, {\bf R}_T) \right] \frac{1}{4} \gamma^+ \nonumber \\
& &+ \frac{1}{4\pi} \left[ \tilde{G}_1 (z, \zeta, {\bf R}_T) \gamma_5 + \gamma_5 \frac{\Rslash_T}{M_h} 
\tilde{H}_1^\sphericalangle (z, \zeta, {\bf R}_T) \right] \frac{1}{4} \gamma^+ \; , 
\label{DeltaPV}
\end{eqnarray}
where $\zeta = (z_1-z_2)/z$. $\tilde{G}_1$ and $\tilde{H}_1^\sphericalangle$ are new P-odd DiFFs. Similarly to what happens in the single-hadron fragmentation~\cite{KK:2011}, they are responsible for a new azimuthally asymmetric term in the cross section for the $e^+ e^- \to (\pi^+ \pi^-) (\pi^+ \pi^-) X$ process, but in a collinear framework. In fact, after averaging over $\cos\theta$ the leading-twist cross section for the semi-inclusive production of collinear pairs reads 
\begin{eqnarray}
& &d\sigma = \frac{3\alpha^2}{4\pi Q^2} \sum_q e_q^2 \Bigg\{ \frac{1+\cos^2\theta_2}{4} \, \left( D_1^q D_1^{\bar{q}} - 
\tilde{G}_1^q \tilde{G}_1^{\bar{q}} \right) \nonumber \\
& &\; + \frac{\sin^2\theta_2}{4} \, \langle \sin\theta \rangle \langle \sin\bar{\theta}\rangle \, \frac{|{\bf R}_T|}{M_h} \frac{|{\bf \bar{R}}_T|}{\bar{M}_h} \cos (\phi_R + \phi_{\bar{R}}) \left( H_1^{\sphericalangle\, q} H_1^{\sphericalangle\,\bar{q}} + \tilde{H}_1^{\sphericalangle\, q} \tilde{H}_1^{\sphericalangle\, \bar{q}} \right) \nonumber \\
& &\; + \frac{\sin^2\theta_2}{4}\, \langle \sin\theta \rangle \langle  \sin\bar{\theta} \rangle \, \frac{|{\bf R}_T|}{M_h} \frac{|{\bf \bar{R}}_T|}{\bar{M}_h} \sin (\phi_R + \phi_{\bar{R}}) \left( \tilde{H}_1^{\sphericalangle\, q} H_1^{\sphericalangle\, \bar{q}} - H_1^{\sphericalangle\, q} \tilde{H}_1^{\sphericalangle\, \bar{q}} \right) \Bigg\} , \, 
\label{e+e-totxsectLM}
\end{eqnarray}
where each DiFF now depends on $(z, M_h^2)$ or $(\bar{z}, \bar{M}_h^2)$, and $\theta_2$ is the angle between the annihilation direction and the $\hat{z}$ axis (see Fig.1 of Ref.~\refcite{noiBelle}). 

In a model calculation based on the spectator approximation, both $H_1^\sphericalangle$ and 
$\tilde{H}_1^\sphericalangle$ have the same size and arise from the interference of pion pairs being produced with relative partial waves with $|\Delta L=1|$. But the former is related to the imaginary part of this interference, while the latter is given by the real part. As such, $\tilde{H}_1^\sphericalangle$ displays a node at $M_h \approx m_\rho$, the mass of the $\rho$ resonance. In Eq.~(\ref{e+e-totxsectLM}), a new azimuthally asymmetric term appears that linearly depends on the P-odd 
$\tilde{H}_1^\sphericalangle$ (or its antiquark partner). The size of the effect should be proportional to 
$( \bar{\theta} / 2 m_\pi)\, ( \tilde{H}_1^\sphericalangle / \tilde{D}_1)^2$. The ratio of P-odd DiFFs is of the same size as that of the P-even partners $H_1^\sphericalangle / D_1$, as mentioned above, and it turns out to be of the order 
0.4~\cite{Bacchetta:2006un}. From Ref.~\refcite{KK:2011}, $\bar{\theta}$ is proportional to the average gradient of the 
$\theta$ field. Using the instanton vacuum model, Kang and Kharzeev estimate it to be approximately equal to 10 MeV. Then, we get $(\bar{\theta}/2m_\pi) (0.4)^2 \approx 0.006$. 


\section*{Acknowledgments}

We acknowledge many useful discussions with D. Boer, who inspired the last section on P-odd fragmentation, and A. Bianconi, who took part in DiFF studies at several steps. This work is partially supported by the Joint Research Activity "Study of Strongly Interacting Matter" (acronym HadronPhysics3, Grant Agreement No. 283286) under the 7th Framework Programme of the European Community. 



\end{document}